\documentclass[twocolumn,english,pra,aps,showpacs]{revtex4-1}
\usepackage[T1]{fontenc}
\usepackage{color}
\usepackage{babel}
\usepackage{mathtools}
\usepackage{bm}
\usepackage{graphicx}
\usepackage{esint}
\usepackage{amsfonts}

\usepackage[usenames,dvipsnames,svgnames,table]{xcolor}
\usepackage[unicode=true,pdfusetitle,
 bookmarks=true,bookmarksnumbered=false,bookmarksopen=false,
 breaklinks=true,pdfborder={0 0 0},backref=false,colorlinks=true]
 {hyperref}
\hypersetup{linkcolor=NavyBlue,urlcolor=NavyBlue,citecolor=NavyBlue}
\usepackage{breakurl}

\begin{document}

\title{Control-enhanced multiparameter quantum estimation}

\author{Jing Liu}
\affiliation{Department of Mechanical and Automation Engineering, The Chinese
University of Hong Kong, Shatin, Hong Kong}

\author{Haidong Yuan}
\affiliation{Department of Mechanical and Automation Engineering, The Chinese
University of Hong Kong, Shatin, Hong Kong}

\begin{abstract}
Most studies in multiparameter estimation assume the dynamics is fixed and focus on identifying the optimal probe
state and the optimal measurements. In practice, however, controls are usually available to alter the dynamics,
which provides another degree of freedom. In this paper we employ optimal control methods, particularly the
gradient ascent pulse engineering (GRAPE), to design optimal controls for the improvement of the precision
limit in multiparameter estimation. We show that the controlled schemes not only capable to provide a higher
precision limit, but also have a higher stability to the inaccuracy of the time point performing the measurements.
This high time stability will benefit the practical metrology where it is hard to perform the measurement at a
very accurate time point due to the response time of the measurement apparatus.
\end{abstract}

%\pacs{03.67.-a, 03.65.Yz, 03.65.-w.}

\maketitle

\section{Introduction}

Quantum metrology has seen a rapid development recently~\cite{Giovannetti2011,GIOV04,cavesprd,rosetta,Fujiwara2008,Paris09,
Escher2011,Demkowicz1,Tsang2013,Rafal2012,durkin2011,Alipour2014,Chaves2013,Correa2015,Toth,Berni,
Xing2016,Tan2013,Tsang2016,LiuJPA2016,LiuNJP2016}, due to its wide applications in quantum imaging~\cite{Shihreview,Boydreview},
quantum sensing~\cite{Degenreview} and quantum Hamiltonian identification~\cite{Shabani2011,Cole2006,
Zhang2014,Wang2016}. In most of these applications there are usually multiple unknown parameters, which falls in the
subject of multiparameter quantum estimation. Due to its generality, multiparameter quantum estimation is gaining increasing
attention. The recent discovery that simultaneous estimation of multiple parameters can provide a
higher precision limit than estimating each parameter individually~\cite{Humphreys2013,Dattareview,Dattamegfield,
Liu2016,Dattagauss,Yue2014,Zhang2017} further prompts the development of this field~\cite{Dattareview}.

Most existing schemes in multiparameter quantum estimation assume the dynamics is fixed, focus on
the identification of the optimal probe states and optimal measurements~\cite{Humphreys2013,Dattareview,
Dattamegfield,Liu2016,Dattagauss,Yue2014,Zhang2017,Yao2014,Yao2014a,Cheng2014,Knott2016,Fan2014,
BerryPRX,Ciampini2016}. In practice, however, additional controls usually can be employed to alter the dynamics
for further improvement of the precision limit. Systematic methods to obtain optimal controls for the improvement
of the precision limit in multiparameter quantum estimation are highly desired.

Controls have been recently employed to improve the precision limit for single-parameter quantum
estimation~\cite{YuanPRL,Liuarxiv,Pang2017}. For multiparameter quantum estimation, systematic methods to
obtain the optimal controls are still lacking, so far optimal controls have only been obtained under unitary
dynamics~\cite{Yuan2016}. In this paper we employ the gradient ascent pulse engineering (GRAPE)~\cite{Khaneja05}
to obtain optimal controls for multiparameter quantum estimation under general Markovian dynamics, this provides
a systematic method on the design of optimally controlled schemes for multiparameter quantum estimation.

GRAPE has been widely applied as the pulse-engineering technique in quantum information,
including the implementation of logic gates for spin systems~\cite{Rowland2012}, Bose-Einstein condensates~\cite{Jager2014}
and nitrogen-vacancy centers~\cite{Wrachtrup2014,Du2015}. GRAPE allows a high flexibility
in the designed pulse shapes and no pulse families need to be assumed in advance~\cite{Khaneja05}.
However, if practical constraints do exist, they can also be easily incorporated into GRAPE~\cite{Dridi2015}.
Thus, GRAPE provides a versatile tool for designing controlled schemes for quantum
parameter estimation~\cite{Liuarxiv}.

Controls can not only improve the sensitivity, but also improve the stability, which is another important factor to
consider in practical quantum metrology~\cite{Degenreview,Durkin2010}. In practice the measurement apparatus usually
has a response time, so the actual time point performing the measurements may be different from the theoretical value.
Thus, the stability of the precision limit to the inaccuracy in the measurement time are important and needs to be
considered. We provide several examples to show the advantages of the controlled schemes and demonstrate that controls
obtained from GRAPE can improve both the precision limit and the stability. Specifically we studied the single-
and two-spin systems with dephasing noises and demonstrated the improvement of the sensitivity and the stability
provided by the controls.

\section{multiparameter estimation} \label{sec:mul}

Typically to estimate multiple parameters encoded in some quantum state $\rho_{\vec{x}}$, where
$\vec{x}=(x_{1},x_{2},...,x_{d})$, one needs to perform a set of positive-operator valued measurements (POVM)
$\{E(y)\}$ ($\sum_{y}E(y)=\openone$ with $\openone$ the identity) on the state, which will give the result $y$ with
probability $p_{y|\vec{x}}=\mathrm{Tr}(\rho_{\vec{x}} E(y))$. From the measurement result one can then construct an estimator
$\vec{\hat{x}}=(\hat{x}_{1},\hat{x}_{2},...,\hat{x}_{d})$. According to the Cram\'er-Rao bound, the covariance of any unbiased
estimator is bounded below by the (classical) Fisher information matrix as~\cite{Helstrom,Holevo}
\begin{equation}
C\geq\mathcal{F}_{\mathrm{cl}}^{-1}. \label{eq:multi_CRB}
\end{equation}
Here $C$ denotes the covariance matrix with the entries $C_{\alpha\beta}:=\sum_{y}(\hat{x}_{\alpha}-x_{\alpha})
(\hat{x}_{\beta}-x_{\beta})p_{y|\vec{x}}$, $\alpha,\beta\in \{1,2,...,d\}$, and $\mathcal{F}_{\mathrm{cl}}$
denotes the classical Fisher information matrix (CFIM) with the $\alpha\beta$th entry given by
\begin{equation}
\mathcal{F}_{\mathrm{cl},\alpha\beta}:=\sum_{y}\frac{(\partial_{x_{\alpha}}p_{y|\vec{x}})
(\partial_{x_{\beta}}p_{y|\vec{x}})}{p_{y|\vec{x}}}.
\end{equation}
The covariance matrix can be further bounded by the quantum Fisher information matrix
\begin{equation}
C\geq\mathcal{F}_{\mathrm{cl}}^{-1}\geq\mathcal{F}_{\mathrm{q}}^{-1}, \label{eq:multi_qCRB}
\end{equation}
where $\mathcal{F}_{\mathrm{q}}$ is the quantum Fisher information matrix (QFIM) with the
$\alpha\beta$th entry given by $\mathcal{F}_{\mathrm{q},\alpha\beta}=\frac{1}{2}\mathrm{Tr}(\rho\{L_{\alpha},L_{\beta}\})$.
$L_{\alpha}$ is the symmetric logarithmic derivative for $x_{\alpha}$ which is the solution to the equation
$\partial_{x_{\alpha}}\rho=(\rho L_{\alpha}+L_{\alpha} \rho)/2$.
In multiparameter quantum estimation the bound given by the QFIM
is usually not saturable~\cite{Dattareview,Matsumoto2002,Ragy2016,Pezze2017},
therefore we will focus on the classical Cram\'er-Rao bound. In this paper we will take
$|\delta \vec{\hat{x}}|^2:=\sum^{d}_{i=1}\delta^2 \hat{x}_{i}$
as the figure of merit, from the Cram\'er-Rao bound we have
\begin{equation}
|\delta \vec{\hat{x}}|^2\geq\mathrm{Tr}\mathcal{F}_{\mathrm{cl}}^{-1}. \label{eq:tot_variance}
\end{equation}
Thus the controls should be designed to minimize $\mathrm{Tr}\mathcal{F}_{\mathrm{cl}}^{-1}$.

For the case with only two parameters, Eq.~(\ref{eq:tot_variance}) reduces to
$|\delta \vec{\hat{x}}|^2 \geq F^{-1}_{\mathrm{cl,e}}$~\cite{Liu2016}, where
$F_{\mathrm{cl,e}}=\det \mathcal{F}_{\mathrm{cl}}/\mathrm{Tr}\mathcal{F}_{\mathrm{cl}}$, here $\det (\cdot)$ denotes the determinant.
For the cases with more than two parameters, it is in general difficult to obtain the
analytical expression for $\mathcal{F}^{-1}_{\mathrm{cl}}$ directly, which means
this function is not very friendly to a gradient-based algorithm in general as in each iteration the
gradient needs to be evaluated numerically which is very demanding computationally.
Usually the bound will be further relaxed to ease the computation. Since
$[\mathcal{F}^{-1}_{\mathrm{cl}}]_{\alpha\alpha}\geq 1/\mathcal{F}_{\mathrm{cl},\alpha\alpha}$, we then have
\begin{equation}
\mathrm{Tr}\mathcal{F}^{-1}_{\mathrm{cl}}\geq\sum_{\alpha} \frac{1}{\mathcal{F}_{\mathrm{cl},\alpha\alpha}}.
\end{equation}
We will then also take the following quantity
\begin{equation}
f_{0}(T)=\left(\sum_{\alpha} \frac{1}{\mathcal{F}_{\mathrm{cl},\alpha\alpha}(T)}\right)^{-1}
\end{equation}
as an objective function to be maximized. However, the obtained controls will be eventually evaluated by their effects on the
figure of merit $\mathrm{Tr}\mathcal{F}^{-1}_{\mathrm{cl}}$, which characterizes the precision limit of $\delta \vec{\hat{x}}$.
In many cases this provides efficient algorithms lead to near optimal controls.

\section{Algorithm} \label{sec:grape}

%----------------------------------------------------
\begin{figure}[tp]
\includegraphics[width=8cm]{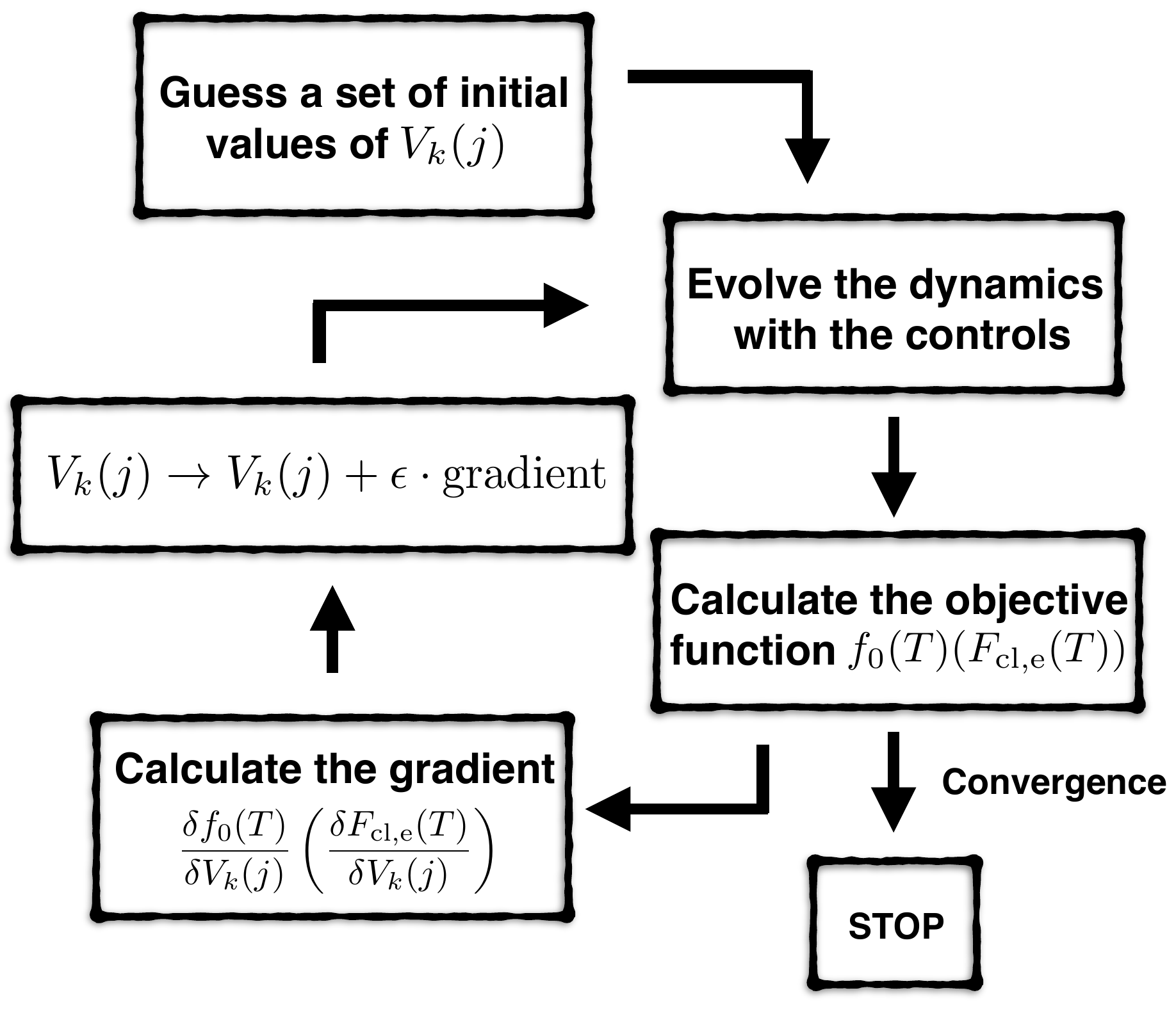}
\caption{(Color online) The flow chart of GRAPE algorithm for multiparameter estimation.}
\label{fig:GRAPE}
\end{figure}
%-----------------------------------------------------

We consider the systems whose dynamics can be described by the master equation
\begin{equation}
\partial_{t}{\rho}(t)=\mathcal{L}[\rho(t)], \label{eq:master_eq}
\end{equation}
where $\mathcal{L}(\rho)=-i[H,\rho]+\Gamma(\rho)$, here
$\Gamma(\rho)$ describes the effect of the noises and $H$ denotes the Hamiltonian. The Hamiltonian can be decomposed as
~\cite{Khaneja05,Ernst1987}
\begin{equation}
H=H_{0}(\vec{x})+\sum_{k=1}^{p}V_{k}(t)H_{k},
\end{equation}
where $H_{0}(\vec{x})$ is the Hamiltonian for the free evolution which contains the unknown parameters $\vec{x}$,
$\sum_{k=1}^{p}V_{k}(t)H_{k}$ is the control Hamiltonian with $V_{k}(t)$ denotes the amplitude
of $k$th control field.

The solution of Eq.~(\ref{eq:master_eq}) at the target time $T$ can be approximated as
$\rho(T)=\Pi_{i=1}^{m}\exp(\Delta t\mathcal{L}_{i})\rho(0)$, with $\mathcal{L}_{i}$ the super-operator
for the $i$th time step and $m=T/\Delta t$ sufficiently large, here the multiplication
in $\rho(T)$ is taken from right to left in the increasing order of $i$.

We then employ gradient ascent pulse engineering (GRAPE)~\cite{Khaneja05} to obtain optimal controls
that can maximize $f_{0}(T)$ ($F_{\mathrm{cl,e}}$ for two-parameter estimation). The flow of the algorithm,
shown in Fig.~\ref{fig:GRAPE}, is as following:
\begin{enumerate}
\item guess a set of initial values for $V_{k}(j)$ ($V_{k}(j)$ is the $k$th control at the $j$th time step);
\item evolve the dynamics with the controls;
\item calculate the objective function $f_{0}(T)$ ($F_{\mathrm{cl,e}}$ for two-parameter estimation);
\item calculate the gradient $\frac{\delta f_{0}(T)}{\delta V_{k}(j)}$
\Big($\frac{\delta F_{\mathrm{cl,e}}}{\delta V_{k}(j)}$ for two-parameter estimation\Big);
\item update $V_{k}(j)$ to $V_{k}(j)+\epsilon \frac{\delta f_{0}(T)}{\delta V_{k}(j)}$;
\item go to step 2 until the objective function converges.
\end{enumerate}
The detailed calculation of the gradient for $f_{0}(T)$, which is based on the gradient
of the entries of CFIM and an essential step for GRAPE, can be found in appendix~\ref{sec:app_gradient}.

GRAPE can be further improved via the quasi-Newton optimization~\cite{Fouquieres2011},
which utilize the gradient history to construct an approximation to the Hessian matrix. Davidon-Fletcher-Powell
formula~\cite{DFP} and Broyden-Fletcher-Goldfarb-Shanno algorithm~\cite{BFGS} are two well applied methods
for this construction. This quasi-Newton optimization can also be utilized here in quantum parameter estimation
for the improvement of GRAPE performance.

Apart from GRAPE, other optimization methods, such as Krotov's
method~\cite{krotovbook,Machnes2011,Reich2012,Goerz2015,Kochreview}, hybrid update method~\cite{Machnes2011} and hybrid optimization
method~\cite{Goerz2015} can all be employed for controlled quantum parameter estimation. The comparison of these methods
for quantum parameter estimation will be studied in future works.

\section{Application} \label{sec:application}
We demonstrate the effect of controls with several examples in Hamiltonian parameter estimation.

\subsection{Noiseless scenario}

\begin{figure}[tp]
\includegraphics[width=8.5cm]{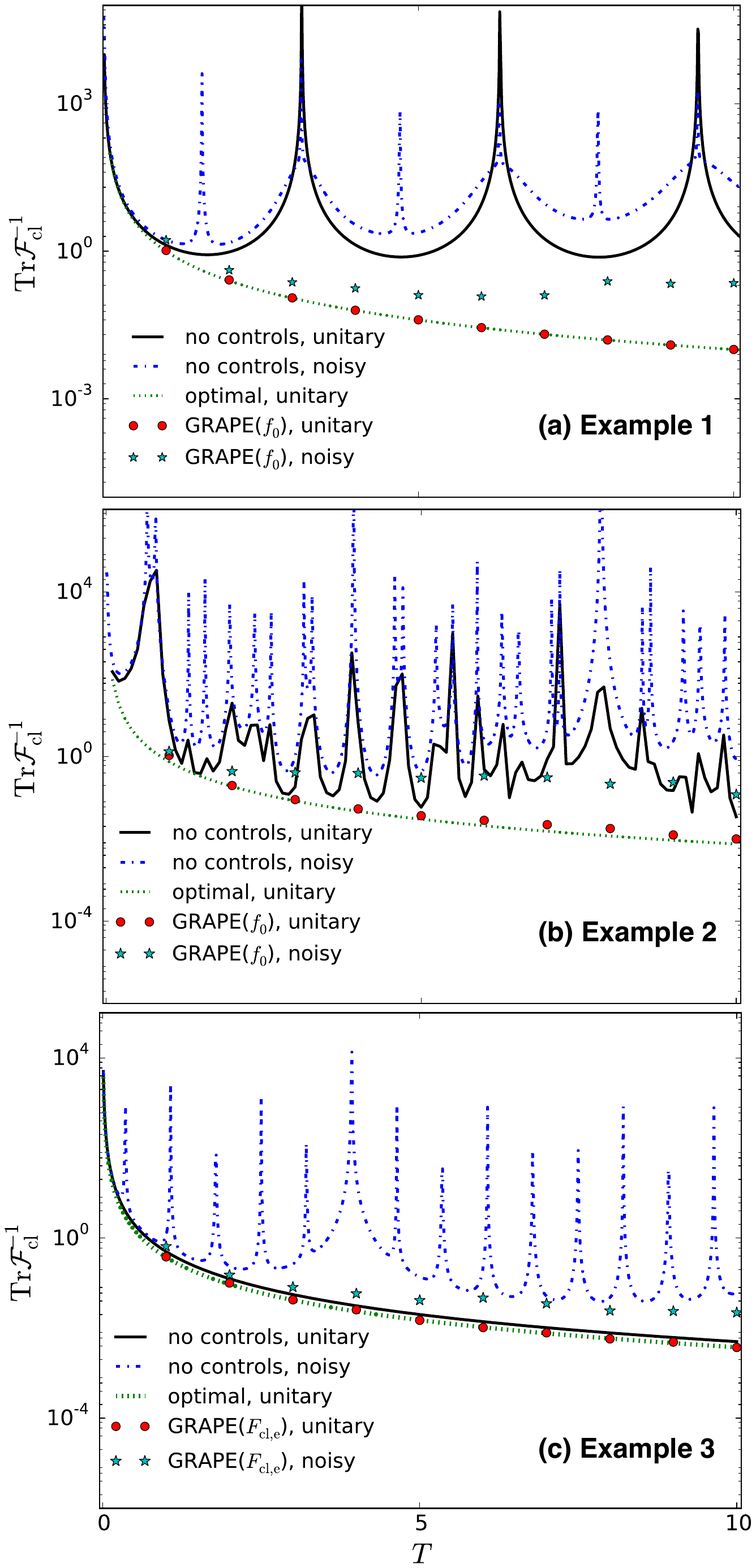}
\caption{(Color online) $\mathrm{Tr}\mathcal{F}^{-1}_{\mathrm{cl}}$ as a function of target time $T$.
The solid black and dashed blue line represent the noiseless and noisy performance of non-controlled
schemes for (a) example 1; (b) example 2 and (c) example 3.The red dots and cyan stars represent noiseless and noisy
performance of controlled schemes in three examples. The controls are obtained via GRAPE with the objective function
$f_{0}(T)$ for (a) and (b) and $F_{\mathrm{cl,e}}$ for (c). The dotted green line represent the optimal total precision
limits. The dephasing rate in (a) is $\gamma=0.2$ and in (b) and (c) is $\gamma_{1}=\gamma_{2}=0.1$. The true values in
(a) are $1,\pi/4,\pi/4$ for $B,\theta,\phi$; in (b) are $1,1.2,0.1$ for $\omega_{1},\omega_{2},g$; in (c) are
$1,1.2$ for $x_{1},x_{2}$.
\label{fig:precision_limit}}
\end{figure}

\emph{Example 1}: Consider a two-qubit system where one qubit is in a magnetic field and the other
qubit acts as an ancillary qubit. The Hamiltonian of this system is $H=H_0+H_\mathrm{c}(t)$ where
\begin{equation}
H_{0}= \vec{B}\cdot\vec{\sigma}^{(1)},
\end{equation}
$\vec{\sigma}^{(1)}=(\sigma^{(1)}_{1},\sigma^{(1)}_{2},\sigma^{(1)}_{3})$ with
$\sigma^{(1)}_{i}=\sigma_{i}\otimes\openone_{2}$,
here $\sigma_{1},\sigma_{2},\sigma_{3}$ are Pauli matrices, $\openone_{2}$ denotes the 2 by 2 identity matrix.
Similarly we will use $\vec{\sigma}^{(2)}=(\sigma^{(2)}_{1},\sigma^{(2)}_{2},\sigma^{(2)}_{3})$
to denote $\sigma^{(2)}_{i}=\openone_{2}\otimes\sigma_{i}$. With the absence of control, the QFIM can be
attained via Bell measurement~\cite{Yuan2016}. For controlled scheme, the local controls on the qubits
\begin{equation}
H_{\mathrm{c}}(t)=\sum_{i=1,2}\vec{V}_{i}(t)\cdot\vec{\sigma}^{(i)}
\label{eq:ctrl_H}
\end{equation}
are employed. This problem has been studied previously and analytical solutions for optimal controls have
been obtained~\cite{Yuan2016}. It has been shown that the optimal controls is to reverse the free evolution
and the optimal initial state is any maximally entangled state, such as
$\frac{1}{\sqrt{2}}\left(|00\rangle\pm|11\rangle\right)$,
and the optimal measurement is the projective measurement on the Bell basis:
$|\Phi^{\pm}\rangle =\frac{1}{\sqrt{2}}\left(|00\rangle\pm|11\rangle\right)$ and
$|\Psi^{\pm}\rangle = \frac{1}{\sqrt{2}}\left(|01\rangle\pm|10\rangle\right)$.
The precision limit for the three parameters $(B,\theta,\phi)$ characterizing
the magnetic field $\vec{B}=(B\sin\theta\cos\phi, B\sin\theta\sin\phi,B\cos\theta)$ under
optimal control is~\cite{Yuan2016}
\begin{equation}
\mathrm{Tr}\mathcal{F}^{-1}_{\mathrm{cl}}=\frac{3}{4T^{2}}.
\label{eq:magfield_noiseless}
\end{equation}
As a confirmation of validity of GRAPE, we compare the performance of the controls obtained from GRAPE
(the objective function is $f_{0}$). It can be seen from Fig.~\ref{fig:precision_limit}(a) that the
performance of the controls (red dots) obtained from GRAPE coincides with the analytical solutions, which
shows that GRAPE is capable to identify the optimal controls in this case.

%-----------------------------------Figure precision limit-------------------------------------
%-------------------------------------------------------------------------------------------------

\emph{Example 2}: Next we consider the Hamiltonian estimation for a two-qubit system
with ZZ coupling, which is a widely used model under strong field. The Hamiltonian in this case is
\begin{equation}
H_{0}=\omega_{1}\sigma^{(1)}_{3}+\omega_{2}\sigma^{(2)}_{3}+g\sigma^{(1)}_{3}\sigma^{(2)}_{3},
\end{equation}
where $\omega_{1}$, $\omega_{2}$ and $g$ are the parameters to be estimated.

We first consider the case without controls. The optimal probe state for this system is
$|++\rangle$ where $|+\rangle=\frac{1}{\sqrt{2}}(|0\rangle+|1\rangle)$ (see detailed derivations in appendix~\ref{sec:appx_HE})
and the corresponding QFIM is $\mathcal{F}_{\mathrm{q}}(T)=4T^{2}\openone_{3}$ ($\openone_{3}$
denotes the 3 by 3 identity matrix). This QFIM is attainable, which is due to the fact that
$\sigma^{(1)}_{3}$, $\sigma^{(2)}_{3}$ and $\sigma^{(1)}_{3}\sigma^{(2)}_{3}$ commute with each other.
The optimal measurement exists~\cite{Humphreys2013,Dattareview} but practically challenging to implement.
In practice, usually the local measurement $E_{\pm}=\{|++\rangle\langle++|,|+-\rangle\langle+-|,|-+\rangle\langle-+|,
|--\rangle\langle--|\}$ is used instead, under this local measurement, the precision (solid
black line in Fig.~\ref{fig:precision_limit}(b)) is typically much worse than the optimal measurement, and it oscillates with time,
which indicates that the time stability of the local measurement is quite poor,
i.e., a small error in the time point chosen to perform the measurement can result very different precisions.

Now if controls are employed, both the precision and the time stability can be improved. This can be seen from
Fig.~\ref{fig:precision_limit}(b), the precision of controlled scheme under the local measurement $E_{\pm}$
(red dots in Fig.~\ref{fig:precision_limit}(b)) essentially coincides with the precision under the optimal
measurement (dotted green line in Fig.~\ref{fig:precision_limit}(b)), it also does not oscillate as in the
non-controlled case. This shows that controls can improve both the precision limit and the time stability.

\emph{Example 3}: In this example we consider the Hamiltonian estimation with XXZ coupling.
The Hamiltonian is
\begin{equation}
H_{0} = -x_{1}\left(\sigma^{(1)}_{1}\sigma^{(2)}_{1}+\sigma^{(1)}_{2}\sigma^{(2)}_{2}\right)
-x_{2}\sigma^{(1)}_{3}\sigma^{(2)}_{3},
\end{equation}
where $x_{1}$ and $x_{2}$ are coupling parameters to be estimated.
First without controls, the optimal QFIM is
\begin{equation}
\mathcal{F}_{\mathrm{q},\mathrm{opt}}=\left(\begin{array}{cc}
8T^{2} & 0\\
0 & 4T^{2}
\end{array}\right),
\end{equation}
which can be attained with the optimal probe state $|\psi_{0,\mathrm{opt}}\rangle=\frac{1}{\sqrt{2}}
|0\rangle(|0\rangle+i|1\rangle)$ (see details in appendix~\ref{sec:xxz}).
The optimal total precision limit for the non-controlled scheme then reads
\begin{equation}
\mathrm{Tr}\mathcal{F}^{-1}_{\mathrm{q,opt}}=\frac{3}{8 T^2}.
\end{equation}
In this case the QFIM is attainable as the generators for $x_{1}$ and $x_{2}$ commute.
However, if we use the more practical local measurement $E_{\pm}$, the CFIM for non-controlled scheme is
$\mathcal{F}_{\mathrm{cl}}=4T^{2}\openone_{2}$, this indicates that the local measurement, $E_{\pm}$,
is only optimal for $x_{2}$. But if we add controls that obtained from GRAPE with $F_{\mathrm{cl,e}}$ as
the objective function, then the precision limit(red dots in Fig.~\ref{fig:precision_limit}(c)) can reach the optimal values
even under the local measurement $E_{\pm}$.

\subsection{Noisy scenario}

We now consider the dynamics with dephasing noises in all examples discussed above.

\emph{Example 1}: Again we consider a two-qubit system where the magnetic field only acts on the first
qubit and the second qubit acts as an ancillary system. For example, in Nitrogen vacancy center, the electron spin can act
as the sensing qubit while a neighboring nuclear spin can act as an ancillary system. Recall that the Hamiltonian is given
by $H=H_0+H_\mathrm{c}(t)$ where $H_{0}= \vec{B}\cdot\vec{\sigma}^{(1)}$ and
$H_{\mathrm{c}}(t)=\sum_{i=1,2}\vec{V}_{i}(t)\cdot\vec{\sigma}^{(i)}$.

Assume that the sensing qubit (first qubit) suffers from the dephasing noise, the dynamics is then described by the master equation
\begin{equation}
\partial_{t}\rho = -i[H,\rho]+\frac{\gamma}{2}\left(\sigma^{(1)}_{3}\rho\sigma^{(1)}_{3}-\rho\right).
\end{equation}
Figure~\ref{fig:precision_limit}(a) (cyan stars) shows the advantages of controlled scheme in improving the precision limit.
From the figure it can also be seen that the controls also suppress the oscillations, thus improve the time stability.

\emph{Example 2}: Consider the Hamiltonian with ZZ coupling $H=H_{0}+H_{\mathrm{c}}(t)$ where
$H_{0}=\omega_{1}\sigma^{(1)}_{3}+\omega_{2}\sigma^{(2)}_{3}+g\sigma^{(1)}_{3}\sigma^{(2)}_{3}$ and
$H_{\mathrm{c}}(t)=\sum_{i=1,2}\vec{V}_{i}(t)\cdot\vec{\sigma}^{(i)}$.
With the presence of dephasing, the dynamics is described by
\begin{equation}
\partial_{t}\rho=-i[H,\rho]+\sum_{i=1,2}\frac{\gamma_{i}}{2}\left(\sigma_{3}^{(i)}\rho\sigma_{3}^{(i)}-\rho\right),
\label{eq:zz_masterequation}
\end{equation}
where $\gamma_{i}$ is the dephasing rate for $i$th qubit. Fig.~\ref{fig:precision_limit}(b) shows that the
controlled scheme improves the precision limit under the local measurement $E_{\pm}$. It can also be seen
that the controls suppressed the oscillations in the precision limit, thus improved the time stability.

\emph{Example 3}: Consider the Hamiltonian estimation with XXZ coupling where
$H_{0} = -x_{1}(\sigma^{(1)}_{1}\sigma^{(2)}_{1}+\sigma^{(1)}_{2}\sigma^{(2)}_{2})
-x_{2}\sigma^{(1)}_{3}\sigma^{(2)}_{3}$ and the control Hamiltonian is
$H_{\mathrm{c}}(t)=\sum_{i=1,2}\vec{V}_{i}(t)\cdot\vec{\sigma}^{(i)}$,
both qubits have the dephasing. The master equation for this system is in the form of Eq.~(\ref{eq:zz_masterequation}),
and we will assume the dephasing rates are the same for both qubits, i.e., $\gamma_{1}=\gamma_{2}=\gamma$.
The CFIM for non-controlled scheme under the practical local measurement
$E_{\pm}$ is given by (see detailed derivation in appendix~\ref{sec:xxz})
\begin{equation}
\mathcal{F}_{\mathrm{cl}}=2T^{2}\left(\begin{array}{cc}
\delta_{+}+\delta_{-} & \delta_{+}-\delta_{-}\\
\delta_{+}-\delta_{-} & \delta_{+}+\delta_{-}
\end{array}\right),
\end{equation}
where
\begin{equation}
\delta_{\pm} := \frac{\cos^{2}\left[2T(x_{1}\pm x_{2})\right]}{e^{2\gamma T}-\sin^{2}\left[2T(x_{1}\pm x_{2})\right]}.
\end{equation}
The precision limit for non-controlled scheme is then
\begin{equation}
\mathrm{Tr}\mathcal{F}_{\mathrm{cl}}^{-1} = \frac{1}{4T^{2}}\left(\frac{1}{\delta_{+}}+\frac{1}{\delta_{-}}\right).
\end{equation}
It can be seen that when $2T(x_{1}\pm x_{2})=\pi/2+n\pi$ ($n=0,1,2...$) this precision blows up, which corresponds to
the peaks in Fig.~\ref{fig:precision_limit}(c).

After adding controls, which are obtained from GRAPE with the objective function $F_{\mathrm{cl,e}}$, the precision
limit (cyan stars in Fig.~\ref{fig:precision_limit}(c)) is improved compared the uncontrolled case
(dashed blue line in Fig.~\ref{fig:precision_limit}(c)). Also, the oscillations is suppressed, indicating
that the controlled scheme provides a higher time stability.

\section{Summary}
In this paper, we employed the optimal control method, GRAPE in particular, in designing optimal controls for
the improvement of the precision limit in multiparameter quantum estimation. Advantages on the precision limit
of controlled schemes are demonstrated through three examples, including the estimation of a
magnetic field, Hamiltonian estimation with ZZ and XXZ couplings. The controls not only improve the precision limit,
but also the time stability, especially for noisy systems.

\begin{acknowledgements}
H.Y. acknowledges partial financial support from RGC of Hong Kong with Grant No. 538213.
J.L. acknowledges partial financial support from the Natural Science Foundation of Zhejiang
Province under Grant No. LY18A050003. We would like
to thank the anonymous referees for the helpful suggestions.
\end{acknowledgements}

%============================================================================
\appendix

\section{Calculation of gradient for the entries of CFIM \label{sec:app_gradient}}

It is known that the entry of CFIM is
\begin{equation}
\mathcal{F}_{\mathrm{cl},\alpha\beta}=\sum_{y}\frac{(\partial_{x_{\alpha}}p_{y|\vec{x}})
(\partial_{x_{\beta}}p_{y|\vec{x}})}{p_{y|\vec{x}}}.
\end{equation}
where $p_{y|\vec{x}}=\mathrm{Tr}(\rho(T)E(y))$. Here $E(y)$
is a POVM measurement which satisfies $\sum_{y}E(y)=\openone$. To
calculate the gradient, we first need to know
\begin{eqnarray}
\frac{\delta p_{y|\vec{x}}}{\delta V_{k}(j)} & = & \mathrm{Tr}\left[\frac{\delta\rho(T)}{\delta V_{k}(j)}E(y)\right]\nonumber \\
 & = & \mathrm{Tr}\left[D_{j+1}^{m}\frac{\delta\rho_{j}}{\delta V_{k}(j)}E(y)\right]\nonumber \\
 & = & -\Delta t\mathrm{Tr}\left[E(y)\mathcal{M}_{j}^{(1)}\right],
\end{eqnarray}
where $\mathcal{M}_{j}^{(1)}=iD_{j+1}^{m}H_{k}^{\times}\rho_{j}$. One then can obtain
\begin{eqnarray}
\frac{\delta\left(\partial_{x_{\alpha}}p_{y|\vec{x}}\right)}{\delta V_{k}(j)}
 & = & -i\Delta t\mathrm{Tr}\left[E(y)\partial_{x_{\alpha}}\!\!\left(D_{j+1}^{m}H_{k}^{\times}\rho_{j}\right)\right] \nonumber \\
 & = & -i\Delta t\mathrm{Tr}\Big\{E(y)\Big[\left(\partial_{x_{\alpha}}D_{j+1}^{m}\right)H_{k}^{\times}\rho_{j} \nonumber \\
 &  & +D_{j+1}^{m}H_{k}^{\times}\partial_{x}\rho_{j}\Big]\Big\}.
\end{eqnarray}
Based on Ref.~\cite{Liuarxiv}, we know
\begin{eqnarray}
\partial_{x_{\alpha}}D_{j+1}^{m} & = & \Delta t\sum_{i=j+1}^{m}D_{i+1}^{m}\left(\partial_{x_{\alpha}}\mathcal{L}_{i}\right)D_{j+1}^{i},\\
\partial_{x_{\alpha}}\rho_{j} & = & \Delta t\sum_{i=1}^{j}D_{i+1}^{j}\left(\partial_{x_{\alpha}}\mathcal{L}_{i}\right)\rho_{i},
\end{eqnarray}
where $\dot{H}^{\times}_{0}=[\partial_{x_{\alpha}}H_{0},\cdot]$. For $j\neq m$,
\begin{eqnarray}
\frac{\delta\left(\partial_{x_{\alpha}} p_{y|\vec{x}}\right)}{\delta V_{k}(j)}
& = & -\Delta^{2}t\mathrm{Tr}\Big[\!\Big(\!E(y)\!\!\sum_{i=j+1}^{m}\!\!D_{i+1}^{m}\dot{H}_{0}^{\times}
D_{j+1}^{i}H_{k}^{\times}\rho_{j} \nonumber \\
&  & +\sum_{i=1}^{j}D_{j+1}^{m}H_{k}^{\times}D_{i+1}^{j}\dot{H}_{0}^{\times}\rho_{i}\Big)\Big].
\end{eqnarray}
for $j=m$, there is
\begin{eqnarray}
\frac{\delta\left(\partial_{x_{\alpha}} p_{y|\vec{x}}\right)}{\delta V_{k}(m)} & = & -i\Delta t\mathrm{Tr}
\left[E(y)H_{k}^{\times}\partial_{x}\rho_{m}\right]\nonumber \\
& = & -\Delta^{2}t\mathrm{Tr}\left[E(y)H_{k}^{\times}\sum_{i=1}^{m}D_{i+1}^{j}\dot{H}_{0}^{\times}\rho_{i}\right]. \nonumber
\end{eqnarray}
Thus, combined above equations, we have
\begin{equation}
\frac{\delta\left(\partial_{x_{\alpha}}p_{y|\vec{x}}\right)}{\delta V_{k}(j)} \nonumber
= -\Delta^{2}t\mathrm{Tr}\left\{ E(y)\left[\mathcal{M}_{j,\alpha}^{(2)}+\mathcal{M}_{j,\alpha}^{(3)}\right]\right\},
\end{equation}
where $\mathcal{M}_{j,\alpha}^{(2)}$ and $\mathcal{M}_{j,\alpha}^{(3)}$ are
\begin{eqnarray}
\mathcal{M}_{j,\alpha}^{(2)} & = & \sum_{i=1}^{j}\mathcal{D}_{j+1}^{m}H_{k}^{\times}
\mathcal{D}_{i+1}^{j}\dot{H}_{0}^{\times}(\rho_{i}),        \label{eq:M} \\
\mathcal{M}_{j,\alpha}^{(3)} & = & (1-\delta_{jm})\!\!\!\sum_{i=j+1}^{m}
\!\!\mathcal{D}_{i+1}^{m}\dot{H}_{0}^{\times}\mathcal{D}_{j+1}^{i}H_{k}^{\times}(\rho_{j}). \nonumber
\end{eqnarray}
Utilizing above expressions, the gradient for $\mathcal{F}_{\mathrm{cl},\alpha\beta}$ is
\begin{eqnarray}
\frac{\delta \mathcal{F}_{\mathrm{cl},\alpha\beta}(T)}{\delta V_{k}(j)}
&=& \Delta t \mathrm{Tr}\left(\tilde{L}_{2,\alpha\beta}\mathcal{M}_{j}^{(1)}\right) \nonumber \\
& & -\Delta^{2}t\mathrm{Tr}\left[\tilde{L}_{1,\beta}\left(\!\mathcal{M}_{j,\alpha}^{(2)}
\!+\!\mathcal{M}_{j,\alpha}^{(3)}\!\right)\right]  \nonumber \\
& & -\Delta^{2}t\mathrm{Tr}\left[\tilde{L}_{1,\alpha}\left(\!\mathcal{M}_{j,\beta}^{(2)}
\!+\!\mathcal{M}_{j,\beta}^{(3)}\!\right)\right],
\label{eq:CFIM_gradient}
\end{eqnarray}
where $\tilde{L}_{1,\alpha(\beta)}$ and $\tilde{L}_{2,\alpha\beta}$ are
\begin{eqnarray}
\tilde{L}_{1,\alpha(\beta)} &=& \sum_{y} \left(\partial_{x_{\alpha(\beta)}}\ln p_{y|\vec{x}}\right) E(y), \\
\tilde{L}_{2,\alpha\beta} &=& \sum_{y} \left(\partial_{x_{\alpha}}\ln p_{y|\vec{x}}\right)
\left(\partial_{x_{\beta}}\ln p_{y|\vec{x}}\right)E(y).
\end{eqnarray}

For two parameter estimation, the objective function is
\begin{equation}
F_{\mathrm{cl,e}}=\frac{\det \mathcal{F}_{\mathrm{cl}}}{\mathrm{Tr}\mathcal{F}_{\mathrm{cl}}}
=\frac{\mathcal{F}_{\mathrm{cl},\alpha\alpha}\mathcal{F}_{\mathrm{cl},\beta\beta}-
\mathcal{F}^{2}_{\mathrm{cl},\alpha\beta}}{\mathcal{F}_{\mathrm{cl},\alpha\alpha}+\mathcal{F}_{\mathrm{cl},\beta\beta}}.
\end{equation}
The corresponding gradient is
\begin{equation}
\frac{\delta F_{\mathrm{cl,e}}}{\delta V_{k}(j)} = \!\!\sum_{\alpha\neq\beta}\!\!
\frac{\mathcal{F}^{2}_{\mathrm{cl},\beta\beta}+\mathcal{F}^{2}_{\mathrm{cl},\alpha\beta}}
{\mathrm{Tr}^{2}\mathcal{F}_{\mathrm{cl}}}\frac{\delta \mathcal{F}_{\mathrm{cl},\alpha\alpha}}{\delta V_{k}(j)}
-\frac{2\mathcal{F}_{\mathrm{cl},\alpha\beta}}{\mathrm{Tr}\mathcal{F}_{\mathrm{cl}}}
\frac{\delta \mathcal{F}_{\mathrm{cl},\alpha\beta}}{\delta V_{k}(j)}
\end{equation}
For parameter estimation with more than two parameters, recall the objective function is
\begin{equation}
f_{0}(T)=\left(\sum_{\alpha} \frac{1}{\mathcal{F}_{\mathrm{cl},\alpha\alpha}(T)}\right)^{-1}.
\end{equation}
The gradient of $f_{0}(T)$ then reads
\begin{equation}
\frac{\delta f_{0}(T)}{\delta V_{k}(j)} =\sum_{\alpha} \frac{f^{2}_{0}(T)}{\mathcal{F}^{2}_{\mathrm{cl},\alpha\alpha}}
\frac{\delta \mathcal{F}_{\mathrm{cl},\alpha\alpha}}{\delta V_{k}(j)}.
\end{equation}

\section{Calculation of noisy CFIM and QFIM in the estimation of magnetic field \label{sec:appx_magfield}}

In this appendix we show the detailed calculation in the estimation of magnetic field.
Recall that the scenario is a two-qubit system where the magnetic field only acts on the first qubit and the
second qubit acts as an ancillary system, the Hamiltonian is given by $H=H_0+H_\mathrm{c}(t)$ where
\begin{equation}
H_{0}= \vec{B}\cdot\vec{\sigma}^{(1)},
\end{equation}
$\vec{\sigma}^{(1)}=(\sigma^{(1)}_{1},\sigma^{(1)}_{2},\sigma^{(1)}_{3})$ with
$\sigma^{(1)}_{i}=\sigma_{i}\otimes\openone_{2}$,
where $\sigma_{1},\sigma_{2},\sigma_{3}$ are Pauli matrices, $\openone_{2}$ denotes the 2 by 2 identity matrix.
Similarly we denote $\vec{\sigma}^{(2)}=(\sigma^{(2)}_{1},\sigma^{(2)}_{2},\sigma^{(2)}_{3})$
with $\sigma^{(2)}_{i}=\openone_{2}\otimes\sigma_{i}$.

With the initial state $\frac{1}{\sqrt{2}}(|00\rangle+|11\rangle)$, the evolved state without noise can be
straightforwardly obtained in the basis $\{|00\rangle,|01\rangle,|10\rangle,|11\rangle\}$ as below
\begin{equation}
|\psi(T)\rangle=\frac{1}{\sqrt{2}}\left(\begin{array}{c}
\cos(BT)-i\sin(BT)\cos\theta \\
-i\sin(BT)\sin\theta e^{-i\phi} \\
-i\sin(BT)\sin\theta e^{i\phi}\\
\cos(BT)+i\sin(BT)\cos\theta\\
\end{array}\right).
\end{equation}
For the further calculation, we rewrite the corresponding density matrix in block form
$|\psi(T)\rangle\langle\psi(T)|=\frac{1}{2}\left(\begin{array}{cc}
X & Y\\
Y^{\dagger} & Q
\end{array}\right), $
where $X$, $Y$ and $Q$ are 2 by 2 matrices. Recall the master equation involving the noise is
\begin{equation}
\partial_{t}\rho = -i[H,\rho]+\frac{\gamma}{2}\left(\sigma^{(1)}_{3}\rho\sigma^{(1)}_{3}-\rho\right).
\end{equation}
The solution for above equation can be expressed in the following form
\begin{equation}
\rho(T)=\frac{1}{2}\left(\begin{array}{cc}
X & e^{-\gamma T}Y\\
e^{-\gamma T}Y^{\dagger} & Q
\end{array}\right).
\end{equation}

Performing the Bell measurements, the probability distribution is
\begin{eqnarray*}
p_{\Phi^{+}} &=& \frac{1}{2}[(1+e^{-\gamma T})\!\cos^{2}(BT)\!+\!(1-e^{-\gamma T})\!\sin^{2}(BT)\!\cos^{2}\theta], \\
p_{\Phi^{-}} &=& \frac{1}{2}[(1-e^{-\gamma T})\!\cos^{2}(BT)\!+\!(1+e^{-\gamma T})\!\sin^{2}(BT)\!\cos^{2}\theta], \\
p_{\Psi^{+}} &=& \frac{1}{2}\sin^{2}(BT)\sin^{2}\theta[1+e^{-\gamma T}\cos(2\phi)], \\
p_{\Psi^{-}} &=& \frac{1}{2}\sin^{2}(BT)\sin^{2}\theta[1-e^{-\gamma T}\cos(2\phi)].
\end{eqnarray*}
According to the probabilities, the entries of CFIM can then be calculated as below
\begin{equation}
\mathcal{F}_{\mathrm{cl},\phi\phi} = \frac{4e^{-2\gamma T}\sin^{2}(BT)\sin^{2}\theta\sin^{2}(2\phi)}
{1-e^{-2\gamma T}\cos^{2}(2\phi)},
\end{equation}
$\mathcal{F}_{\mathrm{cl},\theta\phi} = \mathcal{F}_{\mathrm{cl},B\phi} = 0$ and
\begin{widetext}
\begin{eqnarray*}
\mathcal{F}_{\mathrm{cl},\theta\theta}&=&4\sin^{2}(BT)\left\{\cos^{2}\theta+\frac{\sin^{2}(BT)\sin^{2}\theta\cos^{2}
\theta[(1+3e^{-2\gamma T})\cos^{2}(BT)
+(1-e^{-2\gamma T})\sin^{2}(BT)\cos^{2}\theta]}{[\cos^{2}(BT)+\sin^{2}(BT)\cos^{2}\theta]^{2}-e^{-2\gamma T}
[\cos^{2}(BT)-\sin^{2}(BT)\cos^{2}\theta]^{2}} \right\}, \\
\mathcal{F}_{\mathrm{cl},BB}&=&4T^{2}\cos^{2}(BT)\Big\{\sin^{2}\theta + \sin^{2}(BT)\times\\
& &\frac{[\sin^{4}\theta+e^{-2\gamma T}(1+\cos^{2}\theta)^{2}]
[1-\sin^{2}(BT)\sin^{2}\theta]+2e^{-2\gamma T}\sin^{2}\theta(1+\cos^{2}\theta)[-\cos^{2}(BT)+\sin^{2}(BT)\cos^{2}\theta]}
{[\cos^{2}(BT)+\sin^{2}(BT)\cos^{2}\theta]^{2}-e^{-2\gamma T}[\cos^{2}(BT)-\sin^{2}(BT)\cos^{2}\theta]^{2}} \Big\}, \\
\mathcal{F}_{\mathrm{cl},B\theta}&=&T\sin(2BT)\sin(2\theta)\Big\{1+\sin^{2}(BT) \times \\
& & \frac{(1+e^{-2\gamma T})\cos^{2}(BT)\sin^{2}\theta+(1-e^{-2\gamma T})\sin^{2}(BT)\cos^{2}\theta\sin^{2}\theta-2e^{-2 \gamma T}\cos^{2}(BT)(1+
\cos^{2}\theta)}{[\cos^{2}(BT)+\sin^{2}(BT)\cos^{2}\theta]^{2}-e^{-2\gamma T}[\cos^{2}(BT)-\sin^{2}(BT)\cos^{2}\theta]^{2}} \Big\}.
\end{eqnarray*}
\end{widetext}
Although the CFIM has analytical solution, it is still difficult to obtain the general solution of
$\mathrm{Tr}\mathcal{F}^{-1}_{\mathrm{cl}}$ analytically.
Here we consider the case that $\theta$ is small. The expression of $\mathrm{Tr}\mathcal{F}^{-1}_{\mathrm{cl}}$ approximates to
\begin{eqnarray}
\mathrm{Tr}\mathcal{F}^{-1}_{\mathrm{cl}}&\approx& \frac{e^{2\gamma T}-\cos^{2}(2\phi)}
{2[1-\cos(2BT)]\sin^{2}(2\phi)}\theta^{-2}- \frac{1+\cos(2BT)}{1-\cos(4BT)} \nonumber \\
& &- \frac{1-2e^{2\gamma T}+\cos(4BT)}{4T^{2}[1-\cos(4BT)]}.
\end{eqnarray}

The non-zero eigenvalues of the noisy density matrix are $\lambda_{\pm}=\frac{1}{2}(1\pm e^{-\gamma T})$, which
are independent of the magnetic field. The eigenstates for $\lambda_{\pm}$ are $|\lambda_{+}\rangle=|\psi(T)\rangle$ and
\begin{equation}
|\lambda_{-}\rangle=\frac{1}{\sqrt{2}}\left(\begin{array}{c}
\cos(BT)-i\sin(BT)\cos\theta\\
-i\sin(BT)\sin\theta e^{-i\phi}\\
i\sin(BT)\sin\theta e^{i\phi}\\
\cos(BT)+i\sin(BT)\cos\theta
\end{array}\right).
\end{equation}
It is known that the QFIM can be calculated via the density matrix's non-zero eigenvalues and
corresponding eigenstates~\cite{Liu2014,Liu2014a}. Utilizing above eigenstates, The entries of
QFIM can be calculated as below
\begin{equation}
\mathcal{F}_{\mathrm{q},BB}=4T^{2}\left(\cos^{2}\theta e^{-2\gamma T}+\sin^{2}\theta\right),
\end{equation}
and
\begin{eqnarray}
\mathcal{F}_{\mathrm{q},\theta\theta} &= & 4\sin^{2}(BT)\Big\{\cos^{2}\theta \nonumber \\
& & +\sin^{2}\theta\left[e^{-2\gamma T}\cos^{2}(BT)+\sin^{2}(BT)\right]\!\!\Big\}\!,
\end{eqnarray}
and
\begin{equation}
\mathcal{F}_{\mathrm{q},\phi\phi}\!=\!4\sin^{2}\!\theta\sin^{2}(BT)\!\left[1\!-\!(1-e^{-2\gamma T})
\sin^{2}\!\theta\sin^{2}(BT)\right]\!.
\end{equation}
The off-diagonal entries are
\begin{equation}
\mathcal{F}_{\mathrm{q},B\theta}=(1-e^{-2\gamma T})T\sin(2BT)\sin(2\theta),
\end{equation}
and
\begin{equation}
\mathcal{F}_{\mathrm{q},B\phi}=-2(1-e^{-2\gamma T})T\sin(2\theta)\sin\theta\sin^{2}(BT),
\end{equation}
and
\begin{equation}
\mathcal{F}_{\mathrm{q},\theta\phi}=2(1-e^{-2\gamma T})\sin^{3}\theta\sin(2BT)\sin^{2}(BT).
\end{equation}
it can be seen that the Bell measurement fails to be optimal when noise exists.

\section{Detailed calculation in Hamiltonian estimation with ZZ coupling \label{sec:appx_HE}}

In this appendix we show the detailed calculation in the hamiltonian estimation with ZZ coupling.
The corresponding Hamiltonian is in the form
\begin{equation}
H_{0}=\omega_{1}\sigma^{(1)}_{3}+\omega_{2}\sigma^{(2)}_{3}+g\sigma^{(1)}_{3}\sigma^{(2)}_{3},
\end{equation}
where $\omega_{1}$, $\omega_{2}$ and $g$ are the parameters to be estimated.
It is known that for pure states, the element of QFIM is
\begin{eqnarray}
\mathcal{F}_{\mathrm{q},\alpha\beta}&=& 4\mathrm{Re}\big(\langle\partial_{\alpha}\psi(T)|\partial_{\beta}\psi(T)\rangle \nonumber \\
& & -\langle\partial_{\alpha}\psi(T)|\psi(T)\rangle\langle\psi(T)|\partial_{\beta}\psi(T)\rangle\big).
\end{eqnarray}
In this case, the evolved state $|\psi(T)\rangle=e^{-iHT}|\psi_{0}\rangle$. Utilizing this form, one can have the
diagonal entries of the QFIM as below,
\begin{eqnarray}
\mathcal{F}_{\mathrm{q},\omega_{1}\omega_{2}} & = & 4T^{2}\left(1-\langle\psi_{0}|\sigma^{(1)}_{3}|\psi_{0}\rangle^{2}\right),\\
F_{\mathrm{q},\omega_{2}\omega_{2}} & = & 4T^{2}\left(1-\langle\psi_{0}|\sigma^{(2)}_{3}|\psi_{0}\rangle^{2}\right),\\
F_{\mathrm{q},gg} & = & 4T^{2}\left(1-\langle\psi_{0}|\sigma^{(1)}_{3}\sigma^{(2)}_{3}|\psi_{0}\rangle^{2}\right).
\end{eqnarray}
The off-diagonal entries are
\begin{eqnarray*}
\mathcal{F}_{\mathrm{q},\omega_{1}\omega_{2}} \!& = &\! 4T^{2}\!\!\left(\!\langle\psi_{0}|\sigma^{(1)}_{3}\sigma^{(2)}_{3}|\psi_{0}\rangle
-\langle\psi_{0}|\sigma^{(1)}_{3}|\psi_{0}\rangle\langle\psi_{0}|\sigma^{(2)}_{3}|\psi_{0}\rangle\!\right)\!,\\
\mathcal{F}_{\mathrm{q},\omega_{1}g} \!& = &\! 4T^{2}\!\!\left(\!\langle\psi_{0}|\sigma^{(2)}_{3}|\psi_{0}\rangle
-\langle\psi_{0}|\sigma^{(1)}_{3}|\psi_{0}\rangle\langle\psi_{0}|\sigma^{(1)}_{3}\sigma^{(2)}_{3}|\psi_{0}\rangle\!\right)\!,\\
\mathcal{F}_{\mathrm{q},\omega_{2}g} \!& = &\! 4T^{2}\!\!\left(\!\langle\psi_{0}|\sigma^{(1)}_{3}|\psi_{0}\rangle
-\langle\psi_{0}|\sigma^{(2)}_{3}|\psi_{0}\rangle\langle\psi_{0}|\sigma^{(1)}_{3}\sigma^{(2)}_{3}|\psi_{0}\rangle\!\right)\!.
\end{eqnarray*}

From the expressions of QFIM, one can see that the optimal QFIM is in the form
\begin{equation}
\mathcal{F}_{\mathrm{q,max}}=4T^{2}\openone_{3},
\end{equation}
where $\openone_{3}$ is a 3 by 3 identity matrix. This optimal QFIM can be obtained if the probe state
satisfies the following equations
\begin{eqnarray}
\langle\psi_{0}|\sigma^{(1)}_{3}|\psi_{0}\rangle & = & 0,\\
\langle\psi_{0}|\sigma^{(2)}_{3}|\psi_{0}\rangle & = & 0,\\
\langle\psi_{0}|\sigma^{(1)}_{3}\sigma^{(2)}_{3}|\psi_{0}\rangle & = & 0.
\end{eqnarray}
Denote the probe state as $|\psi_{0}\rangle=a|00\rangle+b|01\rangle+c|10\rangle+d|11\rangle$,
above equations are equivalent to
\begin{eqnarray}
|a|^{2}+|b|^{2}-|c|^{2}-|d|^{2} & = & 0,\\
|a|^{2}+|c|^{2}-|b|^{2}-|d|^{2} & = & 0,\\
|a|^{2}+|d|^{2}-|b|^{2}-|d|^{2} & = & 0.
\end{eqnarray}
Taking into account the normalization $|a|^{2}+|b|^{2}+|c|^{2}+|d|^{2}=1$, the only solution is
$|a|^{2}=|b|^{2}=|c|^{2}=|d|^{2}=1/4.$
Thus, the optimal probe state is in the form
\begin{equation*}
|\psi_{0,\mathrm{opt}}\rangle=\frac{1}{2}\left(|00\rangle+e^{i\phi_{1}}|01\rangle+e^{i\phi_{2}}|10\rangle+e^{i\phi_{3}}|11\rangle\right),
\end{equation*}
where $\phi_{1}$, $\phi_{2}$ and $\phi_{3}$ are relative phases.
The simplest one is $\phi_{1}=\phi_{2}=\phi_{3}=0$, i.e., $|\psi_{0,\mathrm{opt}}\rangle=|++\rangle$,
where $|+\rangle=(|0\rangle+|1\rangle)/\sqrt{2}$.

Now we discuss the noisy case. Consider the dephasing noise for both qubits, the master equation for the system is
\begin{equation}
\partial_{t}\rho=-i[H,\rho]+\sum_{i=1,2}\frac{\gamma_{i}}{2}\left(\sigma_{3}^{(i)}\rho\sigma_{3}^{(i)}-\rho\right).
\end{equation}
The specific solution of $\rho(T)$ is
\begin{widetext}
\begin{equation*}
\left(\begin{array}{cccc}
\rho_{00}(0) & \rho_{01}(0)e^{-i2(g+\omega_{2})T-\gamma_{2}T} & \rho_{02}(0)e^{-i2(g+\omega_{1})T-\gamma_{1}T}
& \rho_{03}(0)e^{-i2(\omega_{1}+\omega_{2})T-(\gamma_{1}+\gamma_{2})T}\\
\rho_{10}(0)e^{i2(g+\omega_{2})T-\gamma_{2}T} & \rho_{11}(0) & \rho_{12}(0)e^{-i2(\omega_{1}-\omega_{2})T-(\gamma_{1}+\gamma_{2})T}
& \rho_{13}(0)e^{i2(g-\omega_{1})T-\gamma_{1}T}\\
\rho_{20}(0)e^{i2(g+\omega_{1})T-\gamma_{1}T} & \rho_{21}(0)e^{i2(\omega_{1}-\omega_{2})T-(\gamma_{1}+\gamma_{2})T}
& \rho_{22}(0) & \rho_{23}(0)e^{i2(g-\omega_{2})T-\gamma_{2}T} \\
\rho_{30}(0)e^{i2(\omega_{1}+\omega_{2})T-(\gamma_{1}+\gamma_{2})T} & \rho_{31}(0)e^{-i2(g-\omega_{1})T-\gamma_{1}T}
& \rho_{32}(0)e^{-i2(g-\omega_{2})T-\gamma_{2}T} & \rho_{33}(0)
\end{array}\right),
\end{equation*}
\end{widetext}
where $\rho_{ij}(0)$ is the initial values of the entries for density matrix. For the probe
state $|++\rangle$, $\rho_{ij}(0)=1/4$ for any $i$ and $j$.
The probability distribution for the measurement $\{|++\rangle\langle++|,|+-\rangle\langle+-|,
|-+\rangle\langle-+|,|--\rangle\langle--|\}$ are
\begin{widetext}
\begin{eqnarray}
p_{++}(T) \!&=&\! \frac{1}{4}\!\left\{1+e^{-\gamma_{1}T}\!\cos(2gT)\cos(2\omega_{1}T)+e^{-\gamma_{2}T}\!\cos(2gT)\cos(2\omega_{2}T)
+e^{-(\gamma_{1}+\gamma_{2})T}\!\cos(2\omega_{1}T)\cos(2\omega_{2}T)\right\}\!,  \\
p_{+-}(T) \!&=&\! \frac{1}{4}\!\left\{1+e^{-\gamma_{1}T}\!\cos(2gT)\cos(2\omega_{1}T)-e^{-\gamma_{2}T}\!\cos(2gT)\cos(2\omega_{2}T)
-e^{-(\gamma_{1}+\gamma_{2})T}\!\cos(2\omega_{1}T)\cos(2\omega_{2}T)\right\}\!,  \\
p_{-+}(T) \!&=&\! \frac{1}{4}\!\left\{1-e^{-\gamma_{1}T}\!\cos(2gT)\cos(2\omega_{1}T)+e^{-\gamma_{2}T}\!\cos(2gT)\cos(2\omega_{2}T)
-e^{-(\gamma_{1}+\gamma_{2})T}\!\cos(2\omega_{1}T)\cos(2\omega_{2}T)\right\}\!,  \\
p_{--}(T) \!&=&\! \frac{1}{4}\!\left\{1-e^{-\gamma_{1}T}\!\cos(2gT)\cos(2\omega_{1}T)-e^{-\gamma_{2}T}\!\cos(2gT)\cos(2\omega_{2}T)
+e^{-(\gamma_{1}+\gamma_{2})T}\!\cos(2\omega_{1}T)\cos(2\omega_{2}T)\right\}\!.
\end{eqnarray}
\end{widetext}

\section{Detailed calculation in Hamiltonian estimation with XXZ coupling\label{sec:xxz}}

In this appendix we show the detailed calculation in the Hamiltonian estimation with XXZ coupling.
Recall the Hamiltonian for the XXZ model is
\begin{equation}
H_{0}=-x_{1}\left(\sigma_{1}^{(1)}\sigma_{1}^{(2)}+\sigma_{2}^{(1)}\sigma_{2}^{(2)}\right)-x_{2}\sigma_{3}^{(1)}\sigma_{3}^{(2)}.
\end{equation}
Because $[\sigma_{3}^{(1)}\sigma_{3}^{(2)},\sigma_{1}^{(1)}\sigma_{1}^{(2)}]=[\sigma_{3}^{(1)}\sigma_{3}^{(2)},\sigma_{2}^{(1)}\sigma_{2}^{(2)}]=0$,
the diagonal entries for QFIM can then be calculated as
\begin{eqnarray}
\mathcal{F}_{\mathrm{q},x_{1}x_{1}} & = & 4T^{2}\Big(2-2\langle\psi_{0}|\sigma_{3}^{(1)}\sigma_{3}^{(2)}|\psi_{0}\rangle \nonumber \\
& & -\langle\psi_{0}|\sigma_{1}^{(1)}\sigma_{1}^{(2)}+\sigma_{2}^{(1)}\sigma_{2}^{(2)}|\psi_{0}\rangle^{2}\Big),
\end{eqnarray}
and $\mathcal{F}_{\mathrm{q},x_{2}x_{2}} = 4T^{2}\left(1-\langle\psi_{0}|\sigma_{3}^{(1)}\sigma_{3}^{(2)}|\psi_{0}\rangle^{2}\right)$.
The off-diagonal entry is
\begin{eqnarray}
\mathcal{F}_{\mathrm{q},x_{1}x_{2}} &=& -4T^{2}\langle\psi_{0}|\sigma_{1}^{(1)}\sigma_{1}^{(2)}
+\sigma_{2}^{(1)}\sigma_{2}^{(2)}|\psi_{0}\rangle \nonumber\\
& & \times \left(1+\langle\psi_{0}|\sigma_{3}^{(1)}\sigma_{3}^{(2)}|\psi_{0}\rangle\right).
\end{eqnarray}
From these expressions, one can find the optimal QFIM is (in the basis $\{x_{1},x_{2}\}$)
\begin{equation}
\mathcal{F}_{\mathrm{q},\mathrm{opt}}=4\left(\begin{array}{cc}
2T^{2} & 0\\
0 & T^{2}
\end{array}\right),
\end{equation}
which can be obtained when the probe state $|\psi_{0}\rangle$ satisfies
\begin{eqnarray}
\langle\psi_{0}|\sigma_{3}^{(1)}\sigma_{3}^{(2)}|\psi_{0}\rangle & = & 0,\\
\langle\psi_{0}|\sigma_{1}^{(1)}\sigma_{1}^{(2)}+\sigma_{2}^{(1)}\sigma_{2}^{(2)}|\psi_{0}\rangle & = & 0.
\end{eqnarray}
Expand the probe state in the computational basis $\{|00\rangle,|01\rangle,|10\rangle,|11\rangle\}$,
i.e., $|\psi_{0}\rangle=a|00\rangle+b|01\rangle+c|10\rangle+d|11\rangle$,
above equations reduce to
\begin{eqnarray}
|a|^{2}-|b|^{2}-|c|^{2}+|d|^{2} & = & 0,\\
b^{*}c+bc^{*} & = & 0.
\end{eqnarray}
Taking into account the normalization relation, above equations can
be reorganized into
\begin{eqnarray}
|a|^{2}+|d|^{2} & = & \frac{1}{2},\\
|b|^{2}+|c|^{2} & = & \frac{1}{2},\\
\cos\left(\phi_{c}-\phi_{b}\right) & = & 0,
\end{eqnarray}
where $\phi_{b}$ and $\phi_{c}$ are the arguments of $b$ and $c$,
respectively. Any probe state satisfies these conditions is the optimal
probe state. One of them is
\begin{equation}
|\psi_{0,\mathrm{opt}}\rangle=\frac{1}{\sqrt{2}}|0\rangle\left(|0\rangle+i|1\rangle\right),
\end{equation}
which will be used as the probe state in the following calculation. The evolved state
can then be further calculated as
\begin{eqnarray}
|\psi_{\mathrm{T},\mathrm{opt}}\rangle
& = & \frac{1}{\sqrt{2}}\left[2\cos(x_{2}T)-e^{-ix_{2}T}\right] |00\rangle \nonumber \\
& & +\frac{i}{\sqrt{2}}e^{-ix_{2}T}\cos\left(2x_{1}T\right)|01\rangle  \nonumber \\
& & -\frac{1}{\sqrt{2}}\sin\left(2x_{1}T\right)e^{-ix_{2}T}|10\rangle.
\end{eqnarray}
Neglecting the global phase $e^{-ix_{2}T}$, this evolved state is equivalent to
\begin{equation}
|\psi_{\mathrm{T},\mathrm{opt}}\rangle \!=\! \frac{1}{\sqrt{2}}\!\left[e^{i2x_{2}T}|00\rangle+\!i\!\cos(2x_{1}T)|01\rangle
\!-\!\sin(2x_{1}T)\!|10\rangle\right]\!.
\end{equation}

Now we perform a practical measurement $\{|++\rangle\langle++|,|+-\rangle\langle+-|,|-+\rangle\langle-+|,|--\rangle\langle--|\}$
and calculate the corresponding CFIM. The probabilities are
\begin{eqnarray}
p_{++} & = & \frac{1}{4}\left\{ 1-\sin\left[2(x_{1}-x_{2})T\right]\right\} ,\\
p_{+-} & = & \frac{1}{4}\left\{ 1-\sin\left[2(x_{1}+x_{2})T\right]\right\} ,\\
p_{-+} & = & \frac{1}{4}\left\{ 1+\sin\left[2(x_{1}+x_{2})T\right]\right\} ,\\
p_{--} & = & \frac{1}{4}\left\{ 1+\sin\left[2(x_{1}-x_{2})T\right]\right\} .
\end{eqnarray}
The corresponding CFIM is (in the basis $\{x_1,x_2\}$)
\begin{equation}
\mathcal{F}_{\mathrm{cl}} = \left(\begin{array}{cc}
4 T^{2 } & 0\\
0 & 4 T^{2}\\
\end{array}\right).
\end{equation}
For $x_{2}$, this measurement is optimal, but for $x_{1}$, it is not. Thus, it is not optimal for joint measurement.

Now we involve the dephasing noise in both qubits. The master equation for the system is
\begin{equation}
\partial_{t}\rho=-i[H,\rho]+\sum_{i=1,2}\frac{\gamma_{i}}{2}\left(\sigma_{3}^{(i)}\rho\sigma_{3}^{(i)}-\rho\right).
\end{equation}
Taking into account the probe state $|\psi_{0,\mathrm{opt}}\rangle$,
the solution for this equation is
\begin{eqnarray}
\rho_{00}(T) & = & \rho_{00}(0)=\frac{1}{2},\\
\rho_{33}(T) & = & \rho_{33}(0)=0,\\
\rho_{03}(T) & = & \rho_{03}(0)e^{-(\gamma_{1}+\gamma_{2})T}=0.
\end{eqnarray}
For $\rho_{01}$ and $\rho_{02}$, the differential equations are
\begin{eqnarray}
\partial_{t}\rho_{01} & = & \left(i2x_{2}-\gamma_{2}\right)\rho_{01}-i2x_{1}\rho_{02},\\
\partial_{t}\rho_{02} & = & \left(i2x_{2}-\gamma_{1}\right)\rho_{02}-i2x_{1}\rho_{01}.
\end{eqnarray}
The solutions are
\begin{eqnarray}
\rho_{01}(T) & = & -\frac{i}{2}e^{\left(i2x_{2}-\gamma_{2}\right)T}\cos\left(2x_{1}T\right),\\
\rho_{02}(T) & = & -\frac{1}{2}e^{\left(i2x_{2}-\gamma_{1}\right)T}\sin\left(2x_{1}T\right).
\end{eqnarray}
For $\rho_{13}$ and $\rho_{23}$, the differential equations are
\begin{eqnarray}
\partial_{t}\rho_{13} & = & -i2x_{2}\rho_{13}-\gamma_{1}\rho_{13}+i2x_{2}\rho_{23},\\
\partial_{t}\rho_{23} & = & i2x_{1}\rho_{13}-i2x_{2}\rho_{23}-\gamma_{2}\rho_{23}.
\end{eqnarray}
Taking into account the initial condition, the solutions are $\rho_{13}(T)=0$
and $\rho_{23}(T)=0$.
For $\rho_{11}$, $\rho_{12}$ and $\rho_{22}$, the differential
equations are
\begin{eqnarray}
\partial_{t}\rho_{12} & = & -\left(\gamma_{1}+\gamma_{2}\right)\rho_{12}-i2x_{1}\left(\rho_{11}-\rho_{22}\right),\\
\partial_{t}\rho_{11} & = & -i2x_{1}\left(\rho_{12}-\rho_{21}\right),\\
\partial_{t}\rho_{22} & = & i2x_{1}\left(\rho_{12}-\rho_{21}\right).
\end{eqnarray}
Taking into account the initial condition the solution for these differential equations are
\begin{equation}
\rho_{12}(T)=-\frac{i}{4}\sin\left(4x_{1}T\right)e^{-(\gamma_{1}+\gamma_{2})T},
\end{equation}
and
\begin{eqnarray}
\rho_{11}(T) & = & \frac{1}{2}-4x_{1}^{2}\frac{1-e^{-(\gamma_{1}+\gamma_{2})T}\cos\left(4x_{1}T\right)}{16x_{1}^{2}
+\left(\gamma_{1}+\gamma_{2}\right)^{2}} \nonumber \\
& &+x_{1}\frac{e^{-(\gamma_{1}+\gamma_{2})T}\left(\gamma_{1}+\gamma_{2}\right)\sin\left(4x_{1}T\right)}{16x_{1}^{2}
+\left(\gamma_{1}+\gamma_{2}\right)^{2}}\!,
\end{eqnarray}
and  $\rho_{22}(T)=1/2-\rho_{11}(T)$.

Next we perform the measurement $\{|++\rangle\langle++|,|+-\rangle\langle+-|,|-+\rangle\langle-+|,|--\rangle\langle--|\}$.
The corresponding probabilities are
\begin{widetext}
\begin{eqnarray}
p_{++} & = & \frac{1}{4}\left[1-\sin\left(2x_{1}T\right)\cos\left(2x_{2}T\right)e^{-\gamma_{1}T}
+\cos\left(2x_{1}T\right)\sin\left(2x_{2}T\right)e^{-\gamma_{2}T}\right],\\
p_{+-} & = & \frac{1}{4}\left[1-\sin\left(2x_{1}T\right)\cos\left(2x_{2}T\right)e^{-\gamma_{1}T}
-\cos\left(2x_{1}T\right)\sin\left(2x_{2}T\right)e^{-\gamma_{2}T}\right],\\
p_{-+} & = & \frac{1}{4}\left[1+\sin\left(2x_{1}T\right)\cos\left(2x_{2}T\right)e^{-\gamma_{1}T}
+\cos\left(2x_{1}T\right)\sin\left(2x_{2}T\right)e^{-\gamma_{2}T}\right],\\
p_{--} & = & \frac{1}{4}\left[1+\sin\left(2x_{1}T\right)\cos\left(2x_{2}T\right)e^{-\gamma_{1}T}
-\cos\left(2x_{1}T\right)\sin\left(2x_{2}T\right)e^{-\gamma_{2}T}\right].
\end{eqnarray}
\end{widetext}
The CFIM for this probability distribution is complicated. Here we consider a simple case
that $\gamma_{1}=\gamma_{2}=\gamma$. With this condition,
\begin{eqnarray*}
\mathcal{F}_{\mathrm{cl},x_{1}x_{1}} & = & \mathcal{F}_{\mathrm{cl},x_{2}x_{2}}=2T^{2}\left(\delta_{+}+\delta_{-}\right),\\
\mathcal{F}_{\mathrm{cl},x_{1}x_{2}} & = & 2T^{2}\left(\delta_{+}-\delta_{-}\right),
\end{eqnarray*}
where the coefficients
\begin{eqnarray}
\delta_{+} & = & \frac{\cos^{2}\left[2T(x_{1}+x_{2})\right]}{e^{2\gamma T}-\sin^{2}\left[2T(x_{1}+x_{2})\right]},\\
\delta_{-} & = & \frac{\cos^{2}\left[2T(x_{1}-x_{2})\right]}{e^{2\gamma T}-\sin^{2}\left[2T(x_{1}-x_{2})\right]}.
\end{eqnarray}
Thus, the total variance is in the form
\begin{equation}
\mathrm{Tr}\mathcal{F}_{\mathrm{cl}}^{-1} = \frac{1}{4T^{2}}\left(\frac{1}{\delta_{+}}+\frac{1}{\delta_{-}}\right).
\end{equation}

%------------------------
%   References
%------------------------

\end{document}